\documentclass[10pt,letterpaper]{article} 
\usepackage{titlesec,float,makeidx,multirow,multicol,graphicx,hyperref,amsmath,amssymb,fancyhdr,changepage,etoolbox,times,enumitem,textcomp,indentfirst,subcaption,caption,setspace}

\usepackage[dvipsnames,svgnames,table]{xcolor}
\usepackage{txfonts}
\usepackage{mathdots}
\usepackage[paperwidth=8.5in,paperheight=11in,top=1in,right=0.5in,bottom=1in,left=0.5in]{geometry}
\usepackage[utf8]{inputenc}
\usepackage[sort&compress,numbers]{natbib} 
\fancypagestyle{firstpage}{
\fancyhead[L,C,R]{}
\fancyfoot[R]{}
\fancyfoot[C]{\thepage}
\fancyfoot[L]{}

}

\RequirePackage{fix-cm}
\newcommand{\size}[2]{{\fontsize{#1}{0}\selectfont#2}}

\titleformat{\section}
  {\normalfont\fontfamily{phv}\bfseries}{\thesection}{10pt}{\MakeUppercase}
\renewcommand*{\thesection}{\fontfamily{phv}\selectfont\textbf{\arabic{section}.}}
\titlespacing{\section}{0pt}{10pt}{0pt}
\titleformat{\subsection}
  {\normalfont\fontfamily{phv}\bfseries}{\thesubsection}{3pt}{}
\renewcommand*{\thesubsection}{\fontfamily{phv}\selectfont\textbf{\arabic{section}.\arabic{subsection}}}
\titlespacing{\subsection}{0pt}{10pt}{0pt}

\DeclareCaptionFormat{upper}{#1#2\uppercase{#3}\par}
\newcommand{\startsquarepar}{%
    \par\begingroup \parfillskip 0pt \relax}
\newcommand{\stopsquarepar}{%
    \par\endgroup}

\hypersetup{
	colorlinks=true,
	linkcolor=black,
	filecolor=blue,      
	urlcolor=blue,
	citecolor=black,
}
\usepackage{nomencl}
\usepackage{bm}
\usepackage{xstring}
\usepackage{xpatch}
\patchcmd{\thenomenclature}
  {\leftmargin\labelwidth}
  {\leftmargin\labelwidth\itemindent 0.25in }
  {}{}

\makenomenclature

\begin{document}
\newgeometry{top=0.5in,right=0.5in,bottom=1in,left=0.5in}
\begin{flushright}
\fontfamily{phv}\selectfont{
\textbf{Conference Paper}\\
\textbf{2022}\\
\textbf{ }\\
\textbf{ }\\[45pt]
\textbf{\size{18}{ }}\\[38pt]
}
\end{flushright}
\begin{center}
    \fontfamily{phv}\selectfont{\size{11}{\textbf{Numerical Investigation of a Rotating Double Compression Ramp Intake\\[20pt]}}}
\end{center}


    \begin{flushright}
        \fontfamily{phv}\selectfont{\textbf{Lubna Margha$^{1,2}$, Ahmed A. Hamada$^{1}$, Othman Ahmed$^{3}$ Ahmed Eltaweel$^{4}$} \newline
        $^1$Department of Ocean Engineering, Texas A$\&$M University, College Station, TX, 77843, USA \newline
        $^2$Department of Aeronautical and Aerospace Engineering, Cairo University, Giza 12613, Egypt \newline 
        $^3$Department of Mechanical Engineering, Badr University in Cairo, Badr, Cairo, 11829, Egypt \newline 
        $^4$Aerospace Engineering Program, University of Science and Technology - Zewail City, Giza, 12578, Egypt \newline }
    \end{flushright}
    
\begin{multicols*}{2}
\section*{Abstract}
The intakes of air-breathing high-speed flying vehicles produce a large share of the thrust propulsion. Furthermore, the propulsion performance of these engines increases when the single-ramp intake is replaced with a multiple-ramps intake. Many scholars numerically and experimentally studied the high-speed engine performance over static single and multiple compression ramps. However, the transient behavior of the flow during the rotation of the double compression ramp from a single ramp is not fully investigated. The present paper aims to numerically investigate the transient shock reflection phenomenon over a rotating double wedge. The problem will start with a 3-Mach number inviscid flow over a single wedge. Then, a portion of the wedge will be rotated upstream at a quite low trailing Mach number to avoid the significant lag effect in the shock waves system. This idea could be applied in the supersonic intake or extensionally in the hypersonic intake of scramjets with a somehow complex mechanism. Further, the length of the rotating portion of the wedge will be changed three times to study its effect on the shock system. The results show a high gain in the pressure due to the rotation of the wedge. Moreover, the wave angles were larger at the low chord ratio value of $w_2/w_i= 0.25$  than at the high values of $w_2/w_i$ at the same second wedge rotating angle, $\theta_2$, resulting in a higher pressure distribution.

Keywords: Regular reflection; Mach reflection; Moving wedge; Dynamic shock waves; Supersonic flow; Dual solution domain.
\mbox{}
\nomenclature[A]{$L(t)$}{Total wedge stream-wise length}
\nomenclature[A]{$w_i$}{Initial wedge chord}
\nomenclature[A]{$\tau$}{Non-dimensional time}
\nomenclature[A]{$\theta_1$}{Stationary wedge angle}
\nomenclature[A]{$\theta_2(t)$}{Rotating wedge angle}
\nomenclature[A]{$h_2(t)$}{Rotating wedge height}
\nomenclature[A]{$h_1$}{Stationary wedge height}
\nomenclature[A]{$M_{\infty}$}{Free-stream Mach number}
\nomenclature[A]{$M_t$}{Trailing-edge Mach number}
\nomenclature[A]{$\mathrm{MS}$}{Mach stem height}
\nomenclature[A]{$H$}{Half height of computational inflow boundary}
\nomenclature[A]{$e$}{Internal energy}
\nomenclature[A]{$\rho$}{Density}
\nomenclature[A]{$p$}{Pressure}
\nomenclature[B]{$\beta_{p}$}{Incident wave angle at reflection$/$triple point}
\nomenclature[B]{$\beta_{i}$}{Intersection wave angle at interaction point}
\nomenclature[B]{$\beta_{t}$}{Transition wave angle}
\printnomenclature[0.68in]
\section{Introduction}
Many aerospace applications in the supersonic and hypersonic flow regimes, such as engine intakes, scram-jet, isolator ducts, and adjacent rockets, incorporate shock waves' interaction and reflection. That is why these phenomena gained the interest of the scientific community. Thus, the scholars investigated the transition and hysteresis between the Regular Reflection (RR) and Mach Reflection (MR) \cite{ben2007shock, ivanov2001transition, yan2003effect, mouton2007mach}. RR consists of two shock waves, the Incident (I) and Reflected (R) shocks. While the MR occurs when the reflected shock is not able to turn the flow and Mach Stem (MS) appears. Thus, the MR is a three-shock wave configuration. MS appears with increasing the Mach number \cite{ben2002hysteresis, ivanov2001flow}, changing the wedge angle \cite{naidoo2014dynamic, laguarda2020dynamics, margha2021dynamic}, and disturbing the flow using laser energy decomposition \cite{yan2003control,khotyanovsky2004parallel,mouton2008experiments}. The appearance of Mach stem affects the performance of engineering devices by changing static and stagnation pressure distribution. These effects were observed in failures of aero-propulsion engines \cite{srivastava2012interaction, graham1996sr}. A method for controlling the transition from regular reflection to Mach reflection is to dynamically change the inclination angle, $\theta$, of the wedge, or divide the wedge into two portions each with a different angle $\theta$.   

During the last couple of decades, moving/rotating a single wedge angle was used to control the transition between the RR and MR. Felthun and Skews \cite{felthun2004dynamic} rotated the wedge towards the upstream of flow around its leading edge. They found that the rotating rate of the wedge highly affects the transition wave angles. Numerical and experimental studies on a rapidly rotating wedge downstream of the flow were performed by Naidoo and Skews \cite{naidoo2011dynamic}. They found that the dynamic transition from MR to RR happened below Von Neumann criteria. However, the transition from RR to MR occurred outside the steady-state theoretical limits. Goyal et al. \cite{goyal2021dynamic} studied the transition from RR to MR by changing the Mach number and the pivot point at the same strong shock reflection domain. They concluded that the effect of the pivot point location on the transition phenomenon is minor. whereas its effect is serious over the development of MS and the reflection point motion. Another innovative method of moving the wedge was proposed by Margha et al. \cite{margha2021dynamic}. They moved the trailing edge upstream with velocity $V (t)$ preserving the wedge height and allowing the total length of the wedge to be changed by changing the wedge angle. This motion of the wedge was studied at $M_\infty = 3$ and different frequencies $\kappa$. They found that the transition wave angle, $\beta_t$, approaches the theoretical limit of Mach Reflection with subsonic downstream flows (MRs). This was found at a relatively high frequency $\kappa = 2$.

In the last five decades, many investigations were conducted for the shock wave interaction over a double wedge. In mid-1975, Bertin and Hinkle \cite{bertin1975experimental} performed experimental tests that agreed with the theoretical results of the shock wave interaction patterns for the MR case. Ben-Dor and Rayvesky \cite{ben1992interaction} studied the interaction of shock waves with the thermal layer within inviscid flows over both concave and convex double wedges. They concluded that increasing thermal layer temperature increases the height of the triple point. Olejniczk et al. \cite{olejniczak_numerical_1996} studied the inviscid flow over a 2-dimensional one-sided double wedge configuration. They identified four shock wave interactions for that particular configuration. Li and Ben-Dor \cite{li_parametric_1997} formulated an analytical model of inviscid flow over the wedge and quantitatively described the physical mechanism by which sonic throat is created and hence Mach stem height is determined. It was found from the analysis that for the same Mach number, Mach stem heights are only determined by geometrical setup. It was shown that the Mach stem height vanishes and flow transforms from MR to RR exactly what happened at von Neumann transition condition. Li and Ben-Dor \cite{li_analytical_1999} performed analytical studies and experiments over static concave double wedges, investigating the transition process between RR and MR. The analytical tests were performed 2-dimensional while the experimental tests were affected by the 3-dimensional type of wedge. They concluded that the effects of the 3-dimensional Wedge were not dominant within the experiment's setup, and the transition angles in 3-dimensional steady flows were very close to the angles at 2-dimensional steady flows. Further, Ivanov et al. \cite{ivanov2001flow} analyzed the hysteresis phenomenon at the transition using 3-dimensional numerical and experimental configurations. Furthermore, Ben-Dor et al. \cite{ben2003self} evolved their work to study double wedge within the range of $5 \leq M \leq 9$ inviscid flow. They aimed to check the existence of the hysteresis phenomenon over double wedge geometry. In addition, they found that there are oscillations induced by flow itself at different angles. Shoev et al. \cite{shoev_numerical_2017} studied the flow over a double wedge with two cases (low/High enthalpy). The coupling between Navier-Stokes (NS) equations with Direct Simulation Monte Carlo (DSMC) was used to perform the simulations. For the low enthalpy case, a good agreement with the previous literature was found. However, the high enthalpy case had a qualitative agreement with previous experimental studies, with a difference in the quantitative results. 

The present paper was intended to dynamically study the shock wave structure over a rotating double wedge at a free-stream Mach number, $M_\infty$, of 3. The rotation starts from a single ramp case at 3 different positions to create the double compression ramp. The rotation was achieved at a quite low rate to avoid a significant change in the shock waves system. During the transition, the effect of varying the rotating wedge length on the shock structure was investigated. This idea can be applied in the supersonic intake or extensionally in the hypersonic intake of scramjets with a somehow complex mechanism.

\section{Computational Model}
\subsection{Model Description}
\begin{figure}[H]
    \centering
    \includegraphics[width=0.99\linewidth]{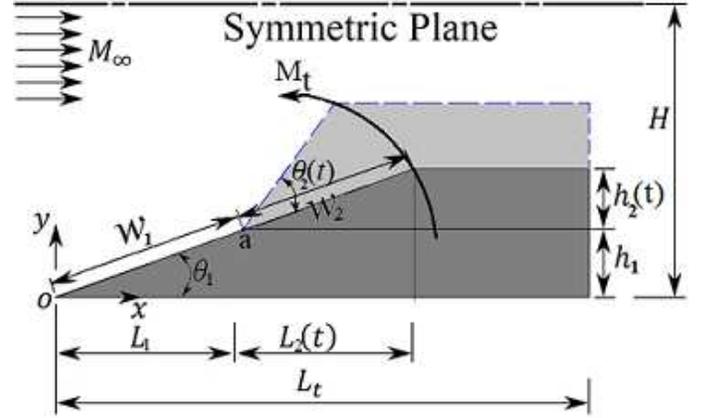}
    \caption{ The geometric schematic of the double rotating wedge.}\label{fig:fig1}
\end{figure}

Figure \ref{fig:fig1} shows the initial single ramp and the rotating double ramps configurations in a supersonic flow with a free-stream Mach number, $M_\infty= 3$. The computational inflow boundary is $2H$ in the transverse-wise and $L_t$ in the stream-wise length. The initial condition is the steady-state solution over a single compression ramp with a deflection angle of $\theta_i=19$ and chord $w_i=1$. Then, a portion of the wedge rotates at pivot point $a$, which is the leading edge of the rotating second wedge. This splits the single ramp to double ramps, one is stationary at the initial wedge angle, $\theta_1=19$. While the second ramp is impulsively rotated with a trailing edge Mach number of $M_t=0.05$. The dynamic shock structure was investigated at three different positions of point $a$. The time-dependent second wedge angle, $\theta_2(t)$, is changed from $19^{\circ}$ to $32^{\circ}$. Both the second ramp's length and height, $L_2(t)$ and $h_2(t)$, change with time step, respectively.The three pivot positions are $a=\frac{1}{4}$, $\frac{1}{2}$, and $\frac{3}{4}$ of the initial chord length, $w_i$. Moreover, Table \ref{table:1} shows the values of the geometric and flow parameters.

\begin{table}[H]
\centering
\caption{System properties and parameters.}
\begin{tabular}{c c} 
 \hline \rule{0mm}{2.5ex}
 Initial wedge's chord, $ w_i = w_1+w_2 $           & $1 m$             \\[0.5ex]
 \hline
 Initial rotating wedge angle, $\theta_2(0)$  & $19^{\circ}$      \\[0.5ex]
 Stationary wedge angle, $\theta_1$           & $19^{\circ}$      \\[0.5ex]
 Final rotating wedge angle, $\theta_2(t_f)$  & $32^{\circ}$      \\[0.5ex]
 Free-stream Mach number, $M_{\infty}$        & $3$               \\[0.5ex]
 Trailing edge Mach number, $M_t$             & $0.05$       \\[0.5ex]
 Total wedge length to the initial chord, $\frac{L_t}{w_i}$ & $1.8$   \\[0.5ex]
 Half domain height to the initial chord, $\frac{H}{w_i}$   & $0.9$   \\[0.5ex]
 Pivot positions to the initial chord, $\frac{w_2}{w_i}$    & $\frac{1}{4}, \frac{1}{2}, \frac{3}{4}$   \\[0.5ex]
\hline
\end{tabular}
\label{table:1}
\end{table}

\begin{figure}[H]
    \centering
    \includegraphics[width=0.86\linewidth]{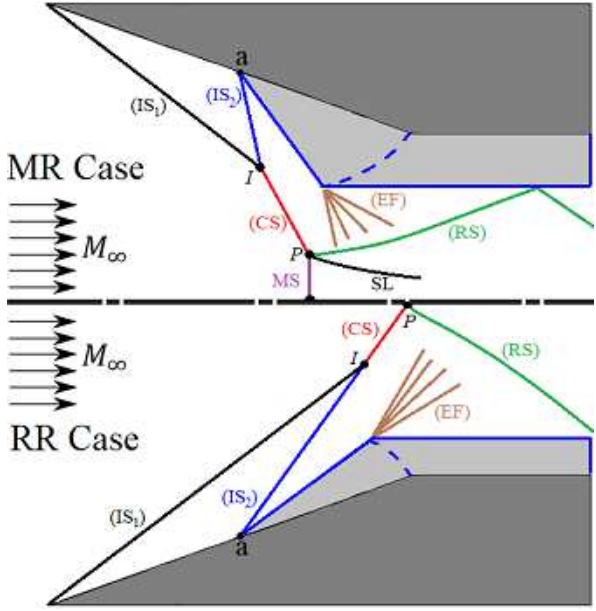}
    \caption{ The RR and MR inflow shock configurations over a double rotating wedge.}\label{fig:fig2}
\end{figure}

The shock-shock wave interaction and the shock wave reflections of the RR and MR shock structures over a rotating slip double wedge are shown in Figure \ref{fig:fig2}. The initial condition starts at a single wedge with a fixed small deflection angle of $\theta_i=\theta_1=19^{\circ}$, in which a RR occurs. The RR shock wave structure is shown in the lower half of the figure, while the MR shock wave configuration is shown in the upper half of the figure. During the rotation of the second wedge, the $1^{st}$ Incident Shock ($IS_1$) wave interacts with the $2^{nd}$ Incident Shock ($IS_2$) at point ($I$). The shock-shock interaction wave angle is measured through the slope between points $I$ and $a$. Then, the two interacting shocks combine forming one combined shock ($CS$). At a relatively small rotating wedge angle, a RR configuration happens when the $CS$ hits the mid-plane of symmetry and reflects regularly at the point $P$, generating a reflected shock wave ($RS$). At higher rotating angles of the second wedge, the transition from RR to MR occurs because the flow is not able to turn to be parallel to the mid-plane. Thus, a normal shock wave with a Mach Stem height (MS) is formed. In this case, the three shock waves gather at the triple point ($P$), where a Slip-Line ($SL$) appears. $\beta_p$ is the combined shock wave angle that is measured from the slope between the reflection/triple point $P$ and the shock interaction point $I$. At the second wedge trailing edge, an Expansion Fan ($EF$) generates and hits the $RS$, deforming it slightly from a straight line. 
\subsection{Governing Equations}
The compressible Euler equations in the Cartesian coordinates are used to compute the supersonic flow over a sharp single and double wedge and are expressed in the conservative form as:	
\begin{equation}
	\frac{\partial Q}{\partial t}+\frac{\partial F}{\partial x}+\frac{\partial G}{\partial y}=0,
\end{equation}
where
\begin{equation}
Q= \begin{bmatrix}
    \rho \\
    \rho u\\
    \rho v\\
    \rho e
\end{bmatrix}, \quad F= \begin{bmatrix}
    \rho u \\
    \rho u^2+p\\
    \rho u v\\
    u (\rho e+p)
    \end{bmatrix}, \quad G= \begin{bmatrix}
    \rho v \\
    \rho u v\\
    \rho v^2+p\\
    v (\rho e+p)
    \end{bmatrix}
\end{equation}
The static pressure is obtained from
\begin{equation}
	p=(\gamma-1) \left( \rho e - \rho \frac{u^2+v^2}{2}\right)
\end{equation}
where $u$ and $v$ are the velocity components in $x$ and $y$ directions, respectively, $p$, $\rho$ and $e$ are pressure, density, and the internal energy of the flow field, respectively, and $\gamma$ is the specific heat ratio of air.
\subsection{Computational Domain}
\begin{figure}[H]
    \centering
    \includegraphics[width=0.99\linewidth]{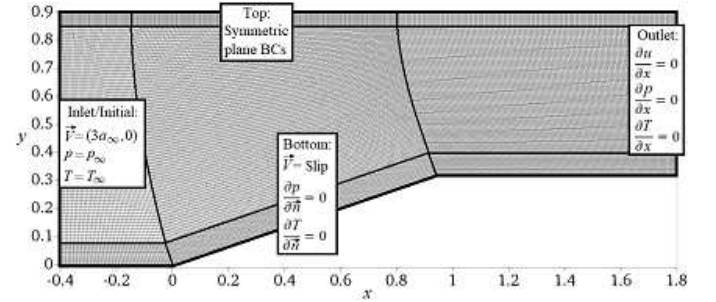}
    \caption{Schematic of the computational domain, the boundary, and initial conditions}\label{fig:fig3}
\end{figure}

The transient compressible flow solver, \textit{rhoCentralDyMFoam}, is used to simulate the flow over the double rotating wedge at $M_\infty=3$. It is a density-based solver implemented in OpenFOAM\textsuperscript{\textregistered}$-$v2006. The letters "\textit{DyM}" in the solver name indicate the ability of the solver to support the dynamic mesh applications. The solver's approach depends on the semi-discrete and upwind-central non-staggered schemes of Kurganov and Tadmor \cite{kurganov2001semidiscrete, greenshields2010implementation}. Half of the computational domain was simulated due to the symmetry of the flow behavior and geometry. Figure \ref{fig:fig3} shows the body-fitted structured mesh for the double wedge at the initial deflection wedge angle of $\theta_i=19^{\circ}$. It also indicates the implemented initial and boundary conditions. 

\begin{figure}[H]
\centering
\includegraphics[width=0.99\linewidth]{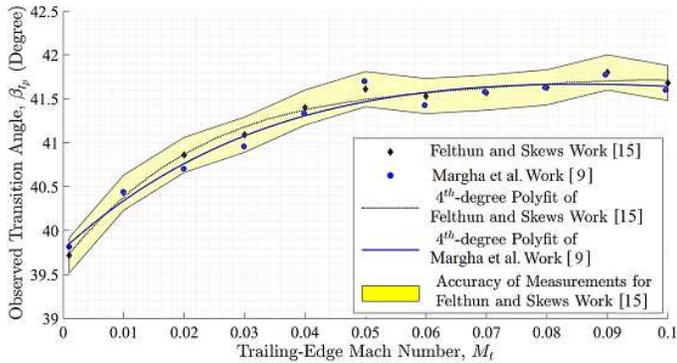}
\caption{Validation with the work of Felthun and Skews [15] by measuring the effect of trailing-edge Mach number on transition angles from regular to Mach reflection [9].}
\label{fig:fig13}
\end{figure}

The mesh independent test on the static and moving wedge was conducted in our previous work \cite{margha2021dynamic} with different mesh sizes at the free-stream Mach number of $3$. Table \ref{table:2} compares mesh sizes by presenting the absolute percentage inaccuracy of the transition wave angle at the reflection point, $\beta_t$, and the Mach stem height, MS, at wedge angle, $\theta=27^{\circ}$. This study resulted in the selection of a mesh size of $2624 \times 720$ cells with an acceptable error percentage of the non-dimensional Mach stem height, MS. The minimum element size in the $x$ and $y$ directions were $0.75mm$ and $0.78mm$, respectively. During the computational time, the time step was adjustable to maintain the Courant–Friedrichs–Lewy (CFL) number at $0.2$. In addition, the verification details of the inserted dynamic code to the \textit{rhoCentralDyMFoam} solver were presented in our previous work \cite{margha2021dynamic}. It was verified with the analysis of Felthun and Skews \cite{felthun2004dynamic} over a dynamic single rotating wedge problem with different rates of the rotation motion, $M_t$ and at the free-stream Mach number of $3$ as shown in Figure \ref{fig:fig13}.  
\begin{table}[H]
\centering
\caption{Independent Grid study: Absolute percentage error of the transition wave angle at the reflection point, $\beta_{t_{p}}$, and the Mach stem height, MS, at wedge angle, $\theta$=$27^\circ$ [9].}
\begin{tabular}{ c c c c c c }
 \hline
 \multirow{2}{*}{Mesh \#} & \multirow{2}{*}{Mesh Size} & \multirow{2}{*}{$\beta_{t_{p}}\ (^{\circ})$} & \multirow{2}{*}{$\frac{\mathrm{MS}}{L(0)} \times 10^{-2}$} & \multicolumn{2}{c}{ $|Error|\%$}\\ & & & &
 $\beta_{t_{p}}$ & $\frac{\mathrm{MS}}{L(0)}$\\ [0.5ex]
 \hline
 1  & $328\ \times 90\ \ $& $41.39$ & 5.29 & $1.22$  &  46.4   \\ 
 2  & $656\ \times 180\ $ & $41.15$ & 7.34 & $0.62$  &  25.6  \\ 
 4  & $1312 \times 360\ $ & $41.03$ & 8.75 & $0.33$  &  11.3  \\
 8  & $2624 \times 720\ $ & $40.96$ & 9.53 & $0.17$  &  3.4   \\
 16 & $5248 \times 1440$  & $40.89$ & 9.87 &   -     &   -    \\[1ex]
 \hline
\end{tabular}
\label{table:2}
\end{table}

\section{Results and Discussion}
The dynamic shock interaction and reflection including the dynamic transition from RR to MR phenomenon was studied over a rotating double wedge. Starting with the steady-state RR shock configuration at a single compression ramp inclined with $\theta_i=19^{\circ}$. Then, the rapid rotation, with a trailing edge Mach number of $M_t=0.05$, began at three different portions ($w_2/w_i=0.25$, $0.5$, and $0.75$) to create the rotating second ramp. While the first ramp was kept fixed at $\theta=19^{\circ}$. Further, the results were compared by rotating the full length of the initial chord (single rotating ramp). The effect of changing the chord of the second ramp relative to the first one was investigated on the dynamic shock structure.

\begin{figure}[H]
    \centering
    \includegraphics[width=0.99\linewidth]{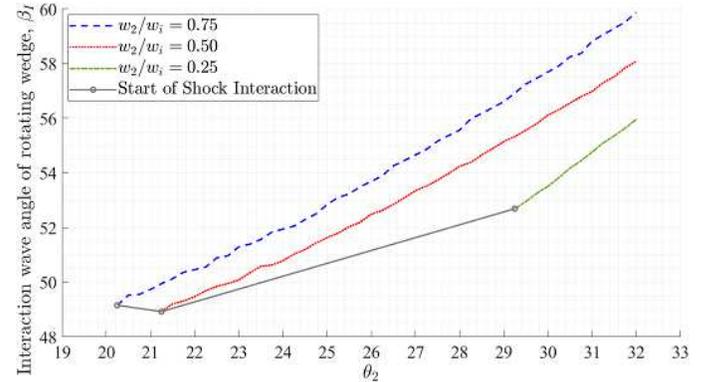}
    \caption{The variation of interaction wave angle of the rotating wedge, $\beta_I^{\circ}$, with the second rotating wedge angle $\theta_2$.} \label{fig:fig7}
\end{figure}

The variation of the shock-shock interaction wave angle, $\beta_I$, with the rotating second ramp deflection angle, $\theta_2$, at $M_t=0.05$, is shown in Figure \ref{fig:fig7}. It is shown that the behavior of the three curves is almost the same. That's because of using the same rotation speed for all tested cases. This is reflected in the same lag effect in the dynamic shock system. Thus, the only effective parameter here is the chord length ratio. The effect of different pivot point locations is shown with different starts interacting wedge and wave angles of the two incident shocks. As the ratio of the second wedge to the initial one, $w_2/w_i$, is large, the start of the interaction point is relatively close to the first wedge's apex, such as the case of $w_2/w_i=0.75$ and $0.5$. On the opposite side, the second wedge angle at the first interaction point was large at a smaller ratio, $w_2/w_i=0.25$. Further, this variation of the wedge angle of the first interaction point is not linear with the variation of pivot point location. 

\begin{figure}[H]
    \centering
    \includegraphics[width=0.99\linewidth]{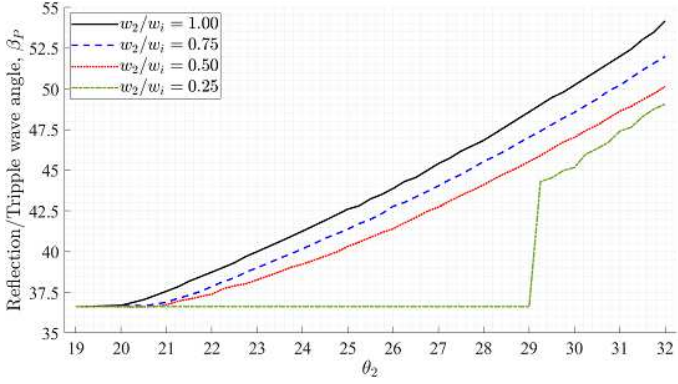}
    \caption{The variation of reflection/triple wave angle, $\beta_P$, with the second rotating wedge angle $\theta_2$. }\label{fig:fig8}
\end{figure}

Moreover, the unsteady wave angle at the reflection/triple point, $\beta_p$, moves upstream with the increase in the rotating ramp angle, $\theta_2$. This variation is compared with the whole length rotating ramp (single rotating ramp), $w_2/w_i=1.0$, as shown in Figure \ref{fig:fig8}. All the simulations started with a steady-state wave angle of $\beta_p=36.6^{\circ}$, corresponding to the initial wedge angle of $\theta_i=19^{\circ}$. For the case of $w_2/w_i=1.0$, the constant value of $\beta_p$ from $\theta_2=19^{\circ}$ to $\theta_2=20^{\circ}$ represents the lag effect due to $M_t=0.05$. Further, the figure shows that the value of $\beta_p$ did not deviate from the initial value with increasing the wedge angle until the start of the shock interaction for $w_2/w_i=0.75$, $0.5$, and $0.25$. The combined shock wave angle is smaller than the second incident wave angle due to the shock-shock interaction phenomenon, which can be observed by comparing the $y$-axes of Figures \ref{fig:fig7} and \ref{fig:fig8}. The deviation in the wave angles for the cases of $w_2/w_i=0.75$ and $0.5$ from the whole rotating wedge, $w_2/w_i=1.0$, is within a half degree to a degree at the start of the rotation (at $\theta=20.5^\circ$ and $\theta=21^\circ$), respectively. Furthermore, this deviation increases with the rotation process. For the case of $w_2/w_i=0.25$, the interaction between the two incident shocks happened at a large second wedge angle of $\theta_2=29^{\circ}$, because it was far from the first wedge's apex. This resulted that the second incident shock from the rotating wedge reflected on the mid-plane of symmetry before interacting with the first incident shock. This interaction caused a sudden jump in the wave angle of the combined shock (\textit{CS}) from $\beta_p=36.6^{\circ}$ to $\beta_p=44.3^{\circ}$.

The sonic and detachment transition criteria were used to measure the dynamic transition from RR to MR for different chord length ratios, $w_2/w_i$. The sonic criterion happens when there is a Mach number of 1 behind the reflected point. After a while, the flow becomes unable to turn parallel to the symmetric plane. At this point, a Mach Stem (MS) generates. This moment is called the detachment point. The values of the transition angles were summarized in Tables \ref{table:3} and \ref{table:4}. The results showed that the difference in the transition angles occurred within a degree between the two cases of $w_2/w_i=0.75$ and $0.5$. Table \ref{table:3} shows that for the case of $w_2/w_i=0.25$, the sonic transition was not possible to be detected. That's because of the sudden interaction between the two incident wave angles at the symmetric plane. This caused the interaction point to be the detachment point, where the MS instantaneously developed, as shown in Figure  \ref{fig:fig9} (j-l). Moreover, the dynamic shock structures of the four different cases of $w_2/w_i$ are shown in Figure \ref{fig:fig9}.

\begin{figure}[H]
    \centering
    \includegraphics[width=0.99\linewidth]{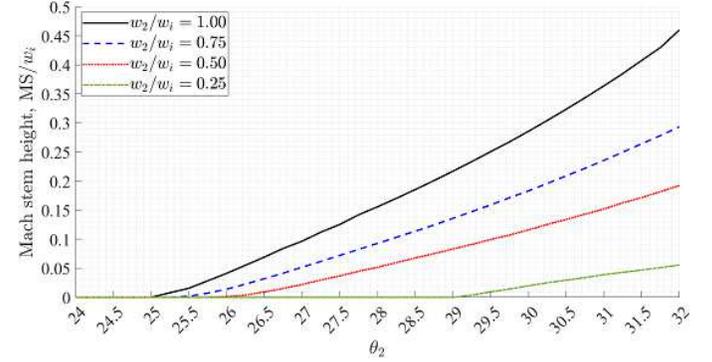}
    \caption{The variation of non-Dimensional Mach stem height with the second wedge angle, $\theta_2^{\circ}$, at different non-dimensional rotating wedge chords $w_2/w_i$ and $M_t=0.05$.} \label{fig:fig6}
\end{figure}

\begin{figure}[H]
	\centering
	\begin{subfigure}{0.3\linewidth}
	    \centering
        \includegraphics[width=0.99\linewidth]{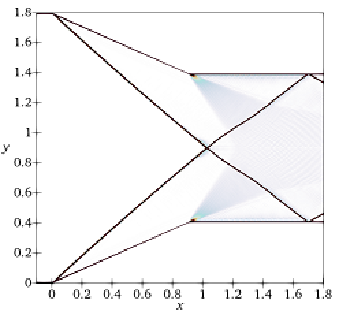}
        \caption{$w_2/w_i=1$ and $\theta_2=24^{\circ}$}\label{fig:fig9_R100_24}
	\end{subfigure}
	\begin{subfigure}{0.3\linewidth}
	    \centering
        \includegraphics[width=0.99\linewidth]{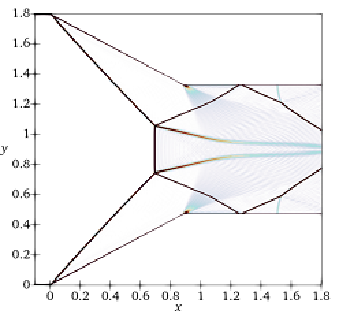}
        \caption{$w_2/w_i=1$ and $\theta_2=28^{\circ}$}\label{fig:fig9_R100_28}
	\end{subfigure}
	\begin{subfigure}{0.3\linewidth}
    	\centering  
        \includegraphics[width=0.99\linewidth]{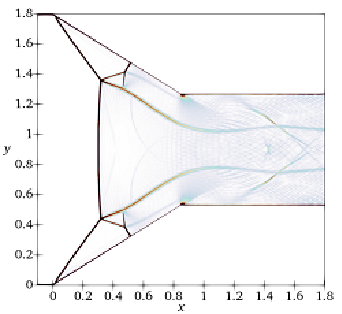}
        \caption{$w_2/w_i=1$ and $\theta_2=32^{\circ}$}\label{fig:fig9_R100_32}
	\end{subfigure}\\
	\begin{subfigure}{0.3\linewidth}
	    \centering
        \includegraphics[width=0.99\linewidth]{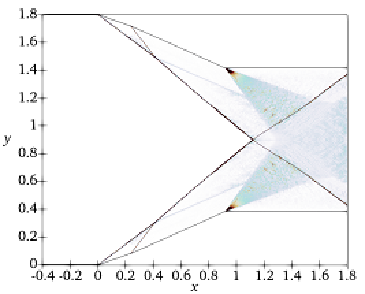}
        \caption{$w_2/w_i=0.75$ and $\theta_2=24^{\circ}$}\label{fig:fig10_R075_24}
	\end{subfigure}
	\begin{subfigure}{0.3\linewidth}
	    \centering
        \includegraphics[width=0.99\linewidth]{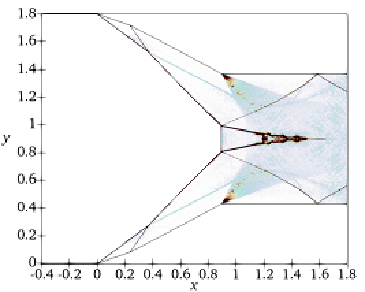}
        \caption{$w_2/w_i=0.75$ and $\theta_2=28^{\circ}$}\label{fig:fig10_R075_28}
	\end{subfigure}
	\begin{subfigure}{0.3\linewidth}
    	\centering  
        \includegraphics[width=0.99\linewidth]{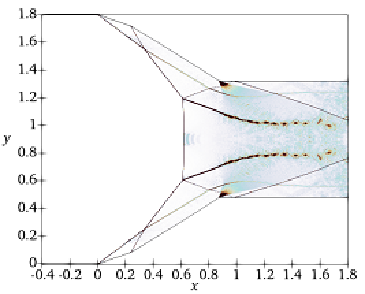}
        \caption{$w_2/w_i=0.75$ and $\theta_2=32^{\circ}$}\label{fig:fig10_R075_32}
	\end{subfigure}\\
	\begin{subfigure}{0.3\linewidth}
	    \centering
        \includegraphics[width=0.99\linewidth]{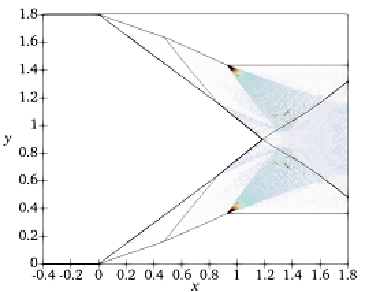}
        \caption{$w_2/w_i=0.5$ and $\theta_2=24^{\circ}$}\label{fig:fig11_R050_24}
	\end{subfigure}
	\begin{subfigure}{0.3\linewidth}
	    \centering
        \includegraphics[width=0.99\linewidth]{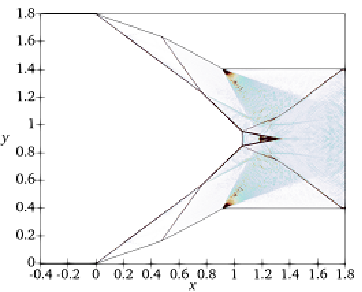}
        \caption{$w_2/w_i=0.5$ and $\theta_2=28^{\circ}$}\label{fig:fig11_R050_28}
	\end{subfigure}
	\begin{subfigure}{0.3\linewidth}
    	\centering  
        \includegraphics[width=0.99\linewidth]{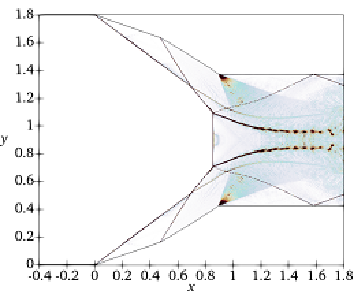}
        \caption{$w_2/w_i=0.5$ and $\theta_2=32^{\circ}$}\label{fig:fig11_R050_32}
	\end{subfigure}\\
	\begin{subfigure}{0.3\linewidth}
	    \centering
        \includegraphics[width=0.99\linewidth]{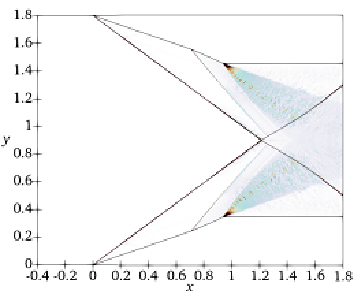}
        \caption{$w_2/w_i=0.25$ and $\theta_2=24^{\circ}$}\label{fig:fig12_R025_24}
	\end{subfigure}
	\begin{subfigure}{0.3\linewidth}
	    \centering
        \includegraphics[width=0.99\linewidth]{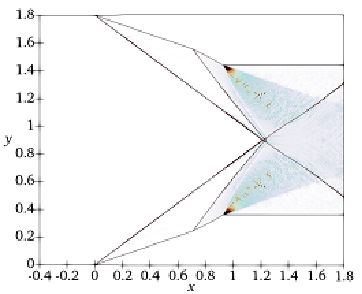}
        \caption{$w_2/w_i=0.25$ and $\theta_2=28^{\circ}$}\label{fig:fig12_R025_28}
	\end{subfigure}
	\begin{subfigure}{0.3\linewidth}
    	\centering  
        \includegraphics[width=0.99\linewidth]{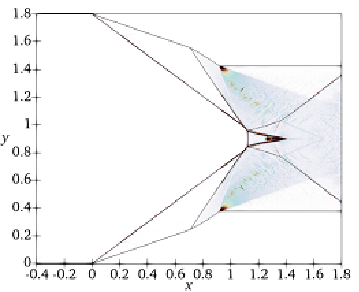}
        \caption{$w_2/w_i=0.25$ and $\theta_2=32^{\circ}$}\label{fig:fig12_R025_32}
	\end{subfigure}
\caption{Normalized velocity gradient while increasing the wedge angle for a single/double rotating wedge at different wedge angles.}\label{fig:fig9}
\end{figure}

\startsquarepar The development of the non-dimensional Mach stem height, MS$/w_i$, with the upstream rotation of the second ramp, $\theta_2$, for \stopsquarepar
\begin{table}[H]
\centering
\caption{Sonic transition angles at different values of $w_2/w_i$ .}
\begin{tabular}{c c c c c } 
 \hline \rule{0mm}{2.5ex}
 $w_2/w_i$ & $\theta_{{t}_2}$ & $\beta_{t_{I}}$ &   $\beta_{t_{P}} $ & $t_t ($ms$)$       \\[0.5ex]
 \hline
    $0.25$  & $-$ &    $-$  &  $-$   &  $-$   \\[0.5ex]
    $0.5$  & $25.047^{\circ}$ &   $51.548^{\circ}$ & $40.362^{\circ}$  & $3.04$    \\[0.5ex]
    $0.75$  & $24.371^{\circ}$  &   $52.254^{\circ}$ & $40.625^{\circ}$ & $4.05$     \\[0.5ex]
 \hline
\end{tabular}
\label{table:3}
\end{table}
\begin{table}[H]
\centering
\caption{Detachment transition angles at different values of $w_2/w_i$ .}
\begin{tabular}{c c c c c } 
 \hline \rule{0mm}{2.5ex}
 $w_2/w_i$ & $\theta_{{t}_2}$ & $\beta_{t_{I}}$ &   $\beta_{t_{P}} $ & $t_t ($ms$)$       \\[0.5ex]
 \hline
    $0.25$  & $29.185^{\circ}$ &    $52.617^{\circ}$  &  $44.55^{\circ}$   &  $2.56$   \\[0.5ex]
    $0.5$  & $26.0^{\circ}$ &   $52.487^{\circ}$ & $41.398^{\circ}$  & $3.52$    \\[0.5ex]
    $0.75$  & $25.432^{\circ}$  &   $53.197^{\circ}$ & $41.91^{\circ}$ & $4.85$     \\[0.5ex]
 \hline
\end{tabular}
\label{table:4}
\end{table}
\noindent different pivot point locations are illustrated in Figure \ref{fig:fig6}. This figure shows the detachment transition criterion from RR to MR when the MS started to appear. The figure shows that the decrease of the rotating wedge chord delayed the MS height, compared with the whole rotating ramp. Further, the rate of the difference in the MS height for each wedge portion case from the case of $w_2/w_i=1.0$ increased during the rotation upstream.

\begin{figure}[H]
	\centering
	\begin{subfigure}{0.99\linewidth}
        \includegraphics[width=0.99\linewidth]{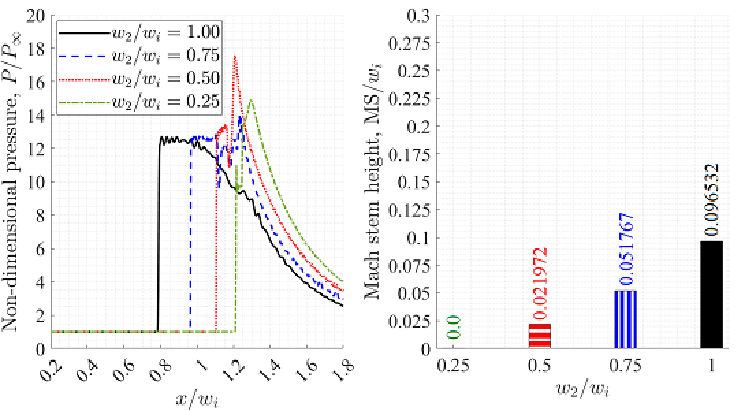}
        \caption{$\theta_2=27^{\circ}$}\label{fig:fig5_2}
	\end{subfigure}
	\begin{subfigure}{0.99\linewidth}
        \includegraphics[width=0.99\linewidth]{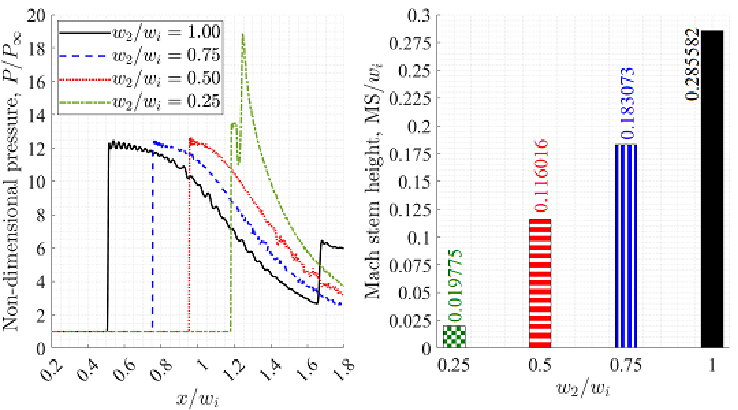}
        \caption{$\theta_2=30^{\circ}$}\label{fig:fig5_1}
	\end{subfigure}
\caption{ The traced pressure distribution through the mid-plane of symmetry at different second-wedge angles, $\theta_2$, and $M_t=0.05$.}\label{fig:fig5}
\end{figure}

In the case of $w_2/w_i=0.25$ and $\theta_2=27^{\circ}$, there is no interaction between the two incident waves. Thus, the MS height was zero, i.e. still RR shock configuration, as shown in the bar graph of Figure \ref{fig:fig5} (a). The reflection points of the two shocks on the symmetric plane are shown in the double jump in the pressure ratio at $x/w_i = 1.215$ and $1.30$, as shown in the left part of Figure \ref{fig:fig5} (a). Further, the strength of the dynamic shock wave from the rotating wedge was higher than the static shock wave, which is indicated by the higher value of the pressure ratio. Thus, the detachment phenomenon happened at the moment of interaction. Additionally, the pressure jump location along the symmetric plane due to shocks was shifted to the right with the decrease of the wedge chord ratio as the apex point of the second ramp was getting closer to the symmetric plane. At each same $\theta_2$, the MS height increased with elongating the rotating wedge portions, as shown in the bar graphs of Figure \ref{fig:fig5}. Despite that, all cases were plotted at the same rotating wedge angle, $\theta_2=30^{\circ}$, and the pressure distribution along the mid-plane of symmetry for the case of $w_2/w_i=0.25$ was higher than the other cases after $x/w_i=1.25$, as shown in Figure \ref{fig:fig5} (b). This indicates that the decrease in the rotating chord ratio increases the combined shock strength (pressure distribution behind).
\end{multicols*}
\begin{multicols*}{2}

\section{Conclusion}
The current research work aims to numerically investigate the effect of changing the rotating wedge chord ratio on the phenomena of the dynamic shock-shock interaction and the dynamic transition from RR to MR. The wedge is divided into two compression ramps where the first one is kept fixed and the second one is rotating with a trailing Mach number of $M_t=0.05$. The pivot point of the rotating wedge was placed at the locations, $a=0.25$, $0.5$, and $0.75$. Further, the results were compared with the dynamics of the whole rotating wedge with the same $M_t$. The study was achieved by measuring the dynamic wave angles, Mach stem height, and pressure distribution along the mid-plane of symmetry. The major conclusions of the current study are:
\begin{itemize}[topsep=1pt,itemsep=1pt,partopsep=1pt, parsep=1pt]
\item The start of the two incident shock-shock interaction was delayed with the decrease of the rotating wedge chord ratio. 
\item The delay in the shock-shock interaction strengthened the combined shock which caused a sudden transition from RR to MR in the case of $w_2/w_i=0.25$. 
\item The dynamic shock systems of the cases $w_2/w_i=0.75$ and $w_2/w_i=0.5$ are relatively close to each other and close to the whole rotating wedge case, unlike the dynamics of the case of $w_2/w_i=0.25$. This was observed in the values of $\beta_I$, $\beta_P$, and MS$/w_i$.
\item At the value of $w_2/w_i= 0.25$, the wave angles were higher than that of the high values of $w_2/w_i$ at the same $\theta_2$, and provided a higher pressure distribution, because the apex point of the rotating wedge was close to the symmetric plane.
\item The variation of the dynamic parameters was not linear with the variation of the wedge chord ratio. Thus, there is a critical wedge chord ratio, where the dynamic shock system becomes very aggressive. A more detailed investigation is required to find the value of this critical ratio. 
\end{itemize}
The future work is to apply the mechanism proposed by Margha et al. \cite{margha2021dynamic} to move the double wedge without changing the throat area and study the dynamic shock system structures and the gain in total pressure.

\bibliography{Margha}
\end{multicols*}
\end{document}